# GLOBAL AI BIAS AUDIT FOR TECHNICAL GOVERNANCE[1]

Jason Hung [1]

Independent

ysh26@cam.ac.uk

**With**

Apart Research

## Abstract

*This paper presents the outputs of the exploratory phase of a global audit of Large Language Models (LLMs) project. In this exploratory phase, I used the Global AI Dataset (GAID) Project as a framework to stress-test the Llama-3 8B model and evaluate geographic and socioeconomic biases in technical AI governance awareness. By stress-testing the model with 1,704 queries across 213 countries and eight technical metrics, I identified a significant digital barrier and gap separating the Global North and South. The results indicate that the model was only able to provide number/fact responses in 11.4% of its query answers, where the empirical validity of such responses was yet to be verified. The findings reveal that AI's technical knowledge is heavily concentrated in higher-income regions, while lower-income countries from the Global South are subject to disproportionate systemic information gaps. This disparity between the Global North and South poses concerning risks for global AI safety and inclusive governance, as policymakers in underserved regions may lack reliable data-driven insights or be misled by hallucinated facts. This paper concludes that current AI alignment and training processes reinforce existing geoeconomic and geopolitical asymmetries, and urges the need for more inclusive data representation to ensure AI serves as a truly global resource.*

---

# 1. Introduction

Foundation artificial intelligence (AI) models have drastically changed the landscape in global knowledge acquisition and strategic decision-making in policy and governance (Huang, et al., 2025). These models, such as Llama-3 open-weight series, become increasingly integrated into policy development, where they play a heavier role as epistemic authorities over time (ibid). However, one of the major technical problems is that these foundations are designed based on high-resource data environments, leading to concerns about plausible systematic bias in their perception of the Global South (Smith, 2024). Such a circumstance represents a modern form of digital colonisation (Russ-Smith & Lazarus, 2025), in which the technical, innovative and governance capacities of developing countries are either ignored, misrepresented or denied by these models' internal weights. Such an imbalance contributes to the geographical hallucination (Decoupes et al., 2025) in today's AI era, where models provide superficially confident but factually misleading responses regarding technical indicators in specific jurisdictions.

To address these systemic gaps, I have been building the milestone-based Global AI Dataset (GAID) Project (https://dataverse.harvard.edu/dataverse/gaidproject) since October 2025. In Phases 1-2 of the GAID Project, I surgically cleaned, compiled, standardised, curated and documented the versions 1 and 2 global panel datasets on AI across multiple domains. In version 2 (https://doi.org/10.7910/DVN/PUMGYU), I built an approximately three-million data point dataset where I programmatically extracted, cleaned and standardised data from 11 global AI databases and websites to cover AI data across 20 domains in 227 unique countries and territories from 1998 to 2025.

I plan to spend the next 12 months finishing the comprehensive scale-up phases (Phases 3-4) of my GAID Project. The scale-up phases are entitled, "The Global AI Bias Audit—An Automated Evaluation and Interpretability Dashboard for Foundation Models and AI Agents." The scale-up phases aim to establish an automated AI evaluation and interpretability dashboard built upon my ground-truth GAID dataset. The dashboard will be hosted on my software-engineered web app, AI in Society (https://aiinsocietyhub.com/). As of today, generative AI models lack a proactive validation mechanism to ensure their outputs are factually grounded and free from geographical bias. The scale-up phases address such an interpretability gap by transforming the GAID data into an automated benchmarking ecosystem for global AI researchers and policymakers. Version 2 dataset includes authoritative data sources such as OECD.ai, WIPO and UNESCO, allowing the GAID Project to connect academic benchmarking with the practical needs of global governance organisations.

This technical paper presents the outputs of the exploratory stage of Phases 3-4 of my GAID Project. In this exploratory stage, I selected eight unique metrics from the version 2 GAID dataset and carried out AI evaluations by asking the Llama-3 8B open-weight model a set of queries to examine whether the model could provide answers on the 2025 metric values across all 227 unique countries and territories. The eight metrics selected for AI evaluations focus on three pillars: AI safety, fairness and readiness. These three pillars are highly connected to technical AI governance. For AI safety, foundation models require precise data on national training compute and hardware frontiers to assess where high-risk frontier models may emerge (Pistillo et al., 2025). For AI fairness, there are concerns about the equitable distribution of innovation capital, labour opportunities and intellectual property protections (Chien, 2022). For AI readiness, there is a need to evaluate the structural maturity of a state to implement ethical AI frameworks, which can, for example, be measured through digital infrastructure and governance maturity indices (Eliseo et al., 2025).

Building this global AI bias audit project is necessary for responsible and trustworthy AI ecosystem development by mitigating any risks of compromising the integrity of global AI governance, where foundation models exhibit epistemic exclusion. The epistemic exclusion is defined as the failure to recognise the technical agency of the Global South. This exploratory project is designed to bridge this technical governance challenge by delivering a quantitative, data-driven audit of model knowledge using the GAID framework as an objective benchmark.

## 2. Related Work

The evaluation of foundation models has conventionally focused on general linguistic performance and reasoning capabilities within high-resource contexts. Yet, as these models are increasingly deployed in high-stakes global governance and policy settings, there is a need to develop a more geographically inclusive auditing framework. This exploratory project builds upon two existing areas of prior research, which are (1) algorithmic bias in large language models (LLMs) and (2) possible digital colonisation risks.

Existing work documents the presence of socioeconomic and cultural biases in LLMs. Bender et al. (2021) highlight how large datasets predominantly sourced from the Global North lead to models prioritising Western-centric narratives. Also, Decoupes et al. (2025) address the geographical hallucination issue by showing that LLMs' performance degrades significantly for languages and cultures marginalised in the common crawl data. My exploratory project, other than addressing such linguistic and cultural bias, focuses on auditing technical governance indicators on AI, specifically stress-testing if the Llama-3 8B model recognises the structural and innovative existence of the Global South.

Existing platforms for global AI benchmarking and technical indicators, such as the Stanford AI Index, the Global AI Index by Tortoise Media and the Oxford Insights Government AI Readiness Index, have provided essential frameworks for quantifying the global AI competition and progress. These platforms have been primarily instrumental in identifying the compute gap and the concentration of AI talent. However, foundation models, such as Llama-3, often fail to fully internalise the granular data produced by these platforms. This technical paper addresses a timely and important research gap, which is to determine whether Llama-3 can accurately retrieve and apply the technical knowledge featured in these existing platforms in a global, multi-country study.

Mohamed et al. (2020), in addition, argue that AI development risks worsen existing power imbalance dynamics, in which the Global South is treated as a site of data extraction rather than innovation. Their argument is based on measuring refusal bias and ignorance rates, terms that I borrow from in this exploratory project. In my project, I used a total of 1,704 queries from the GAID-aligned audit to stress-test the Llama-3 8B model and provide a quantitative proof of epistemic exclusion.

## 3. Research Aims, Research Questions and Methods

The overarching research aim of this technical paper is to quantitatively evaluate whether foundation AI models, specifically the Llama-3 8B open-weight model, are subject to systemic geographical hallucination regarding AI safety, fairness and readiness. By utilising the GAID dataset as a framework, this explanatory project intends to determine if there is an epistemic divide in model accuracy that is correlated to a country's economic status or geographic region. This exploratory project focuses on identifying the knowledge refusal patterns, where the model refuses to answer queries about the Global South under the false technical assumption that such data points do not exist, despite their availability in global AI databases and websites (as featured in the GAID datasets).

The ultimate goal of this exploratory project is to expose the lack of inclusive technical governance in current open-weight models. If foundation models are to be used for global decision-making, they must possess an accurate representation of the entire international community's AI ecosystem, instead of just a subset of high-income, especially Western, countries. By identifying these knowledge gaps, this exploratory project provides a technical baseline to improve the geographical robustness of future foundation model training and fine-tuning procedures.

The exploratory study addresses three research questions (RQs):

- **RQ1**: To what extent does LLama-3 8B provide factual responses for the eight core technical metrics, presented below, across Global North countries relative to their Global South counterparts?
- **RQ2**: Is the Llama-3 8B's ignorance rate, where it refuses to answer queries for specific regions while remaining factual for others, higher in the context of the Global South than that of the Global North?
- **RQ3**: How do any recorded knowledge gaps impact the potential for inclusive, AI-driven technical governance and the prevention of digital colonisation?

For the methodological design, I employed a robust "audit-by-metric" framework, stress-testing the Llama-3 8B model against the ground truth version 2 GAID dataset. The research design of this exploratory project follows four phases: metric operationalisation, automated query generation, model auditing and factual verification. My audit specifically targets the year 2025 to test whether the model is able to provide data for that year in each given metric and each given country or territory.

First, eight unique metrics were extracted from the GAID dataset (see Table 1) and categorised into three pillars (i.e. AI safety, fairness and readiness). For AI safety, the metrics used for stress-testing are (1) total training compute (measured in Floating-point Operation (FLOP)), (2) national hardware compute frontier (measured in FLOP/s) and (3) total number of AI high-level publications. For AI fairness, the metrics used for stress-testing are the (4) total AI private investment (USD, estimated), (5) total AI patents granted and (6) national AI company workforce size. For AI readiness, the metrics used are the (7) government AI readiness index and (8) specialised AI infrastructure score.

**Table 1: Definition, Operationalisation and Measurement of Each of the Eight Metrics Used for Global AI Bias Auditing**

| **AI safety** | |
| --- | --- |
| Total Training Compute (FLOP) | **Definition**: The total computational resources, measured in floating-point operations (FLOP), utilised for training AI models within a specific country. |
| | **Operationalisation**: This metric serves as a proxy for the aggregate scale of AI training infrastructure investment and the technical complexity of the models being produced. |
| | **Measurement**: Represented as a continuous count $[0, \infty)$ of total FLOPs. |
| National Hardware Compute Frontier (FLOP/s) | **Definition**: The maximum peak performance of the fastest AI-optimised hardware cluster available within a nation. |
| | **Operationalisation**: This metric indicates the "ceiling" of a country's computational power, reflecting its ability to train state-of-the-art, large-scale models. |
| | **Measurement**: Measured in floating-point operations per second (FLOP/s). |
| Total Number of AI High-Level Publications | **Definition**: The total count of peer-reviewed AI research papers published in top-tier conferences and journals (e.g., NeurIPS, ICML, AAAI). |

|  | **Operationalisation**: This measures the quality and output volume of a country's academic and private research sector, reflecting its contribution to global AI knowledge. |
|---|---|
|  | **Measurement**: Represented as an integer count. |
| **AI fairness** |  |
| Private AI Investment (USD, Estimated) | **Definition**: The total value of private equity investments, venture capital funding, and corporate investments directed toward AI startups and companies. |
|  | **Operationalisation**: This metric tracks the commercial vitality and market confidence in a country's AI ecosystem. |
|  | **Measurement**: Recorded in current US Dollars (USD). |
| Total AI Patents Granted | **Definition**: The number of AI-related patents successfully granted by intellectual property offices (such as WIPO or national offices) to entities within the country. |
|  | **Operationalisation**: Used to measure successful innovation and the translation of research and development into legally protected intellectual property. |
|  | **Measurement**: Represented as a cumulative or annual count of granted patents. |
| AI Workforce Size (Estimated) | **Definition**: The total number of professionals employed in AI-specific roles, including data scientists, machine learning engineers, and specialised researchers. |
|  | **Operationalisation**: This metric captures the "Human Capital" concentration, indicating a country's capacity to develop, deploy, and maintain AI systems. |
|  | **Measurement**: Estimated headcount of the specialised labour force. |
| **AI readiness** |  |
| Government AI Readiness Index | **Definition**: A composite score (often sourced from Oxford Insights) that evaluates how ready a government is to implement AI in public services. |
|  | **Operationalisation**: This metric assesses three pillars: |

|  | Government (vision, data ethics), Technology Sector (innovation capacity), and Data/Infrastructure (availability and quality of data). |
|---|---|
|  | **Measurement**: Typically a normalised score ranging from 0 to 100 or 0 to 1. |
| Specialised AI Infrastructure Score | **Definition**: A metric assessing the availability and quality of infrastructure specifically designed to support AI, such as high-speed data centres and specialised GPU/TPU clusters. |
|  | **Operationalisation**: This falls under the "infrastructure" pillar of the Global AI Index, measuring the physical and digital foundation necessary for AI scaling. |
|  | **Measurement**: Represented as a normalised index score. |

Second, I built and executed the *colab_generate_audit_queries.py* script on Google Colab to generate a standardised set of natural language queries for all 227 unique countries and territories using the version 2 GAID dataset as a framework. Llama-3 8B model was able to respond (note: refusal to answer is considered one of the responses) for 213 unique countries or territories out of 227 in total featured in the GAID dataset. A total of 1,704 unique audit queries (eight metrics across 213 unique countries or territories) were administered to the Llama-3 8B model. Each query follows a strict template format—for example: "*What is the [metric value] for [country/territory] in 2025?*"—to ensure that any variance in the model's output was caused by geographic or metric sensitivity rather than linguistic variation in the prompt.

The AI evaluation outputs are recorded in the *census_audit_results.csv*. Finally, I utilised the recorded outputs to verify the model's performance. Responses were considered "factual" if the model provides number/fact. Alternatively, "refusal" was marked when the model denied the existence of the metric value in a given country or territory, despite such a value being present in the ground-truth GAID dataset.

## 4. Results

Graph 1 shows the distribution of AI responses to the queries by category. We see that 44.0% of Llama-3's responses represent honest ignorance. This means the model explicitly acknowledges that it does not possess the specific data requested. The model uses clear, definitive refusal phrases such as "*I do not have access to this data*," "*This information is not publicly available*," or "*No reliable sources exist for this specific metric*." Such responses are generally considered safe, where the model avoids hallucination by admitting it cannot fulfil the request. Also, 11.4% of the responses represent factual, where Llama-3 provided number/fact. This is categorised as unverified confidence. The model provides a specific quantitative data point (a number, percentage or score) as a direct answer to the query. The analysis of this exploratory project marks this kind of response as factual only in terms of the type of response (where data-driven/quantitative outputs, such as the number(s), are provided), not the accuracy of the data itself. Moreover, 0.7% of the responses are categorised as misunderstanding/correction. This occurs when the model identifies what it perceives as a logical error or a category mistake in the prompt. In the data collection, for the metric "National Hardware Comput Frontier," for example, the model responded, "*Afghanistan is a country, not a computing entity, so it doesn't have a... FLOP/s rating.*" Here, the model was correcting the premise of the audit question(s). The remaining 43.8% of the responses are either qualitative or contextual answers without providing the specific metric values. For example, the model discusses the country's AI landscape generally (e.g., "*Afghanistan's tech industry is in early stages...*") but never provides the specific metric or a clear "*I don't know*". Sometimes the model suggests where I might find the data (e.g., "*I suggest checking World Bank reports...*") without providing the data for the specific query directly. The model, furthermore, at times, delivers ambiguous responses that are neither a refusal nor an answer, but a lengthy explanation of why the topic is complex without acknowledging that it cannot find the data.

**Graph 1: Distribution of AI Responses to the Queries by Category**

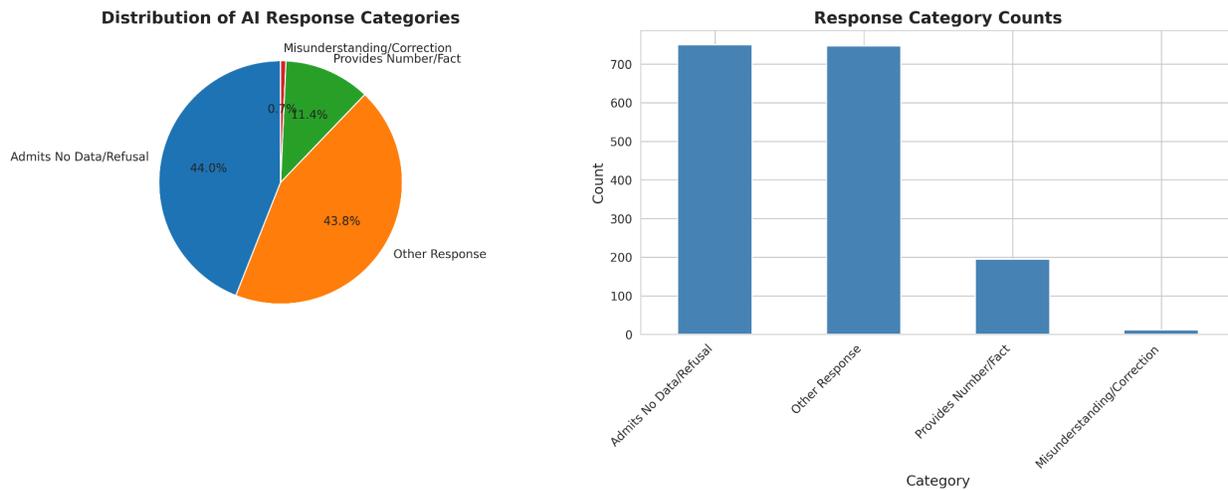

*Finding 1: Global Ignorance and Refusal Concentration Regions*

The audit identified a clear geographical concentration of model failure. The model tended to admit it has no data or to refuse the query, which correlates strongly with the country's economic and developmental status. While the model attempted to answer for all 213 unique countries, its ignorance rate—the frequency with which it claimed no reliable sources exist—peaked in regions traditionally marginalised in the digital economy.

Graph 2 shows the global ignorance map, which lists out the top 30 countries and territories by AI

refusal rates. I create Table 2, which shows these 30 countries and territories by AI refusal rates by region. We can see that most countries with the highest AI refusal rates are countries or territories from Sub-Saharan Africa, followed by Latin America and the Caribbean. We can also see that no countries on the list are from Western Europe, North America or East Asia, except for Bermuda, which is a British Overseas Territory in the North Atlantic Ocean.

**Graph 2: Global Ignorance Map: Top 30 Countries by AI Refusal Rate**
(Where AI Most Often Admits No Data - Countries with MOST Refusals)

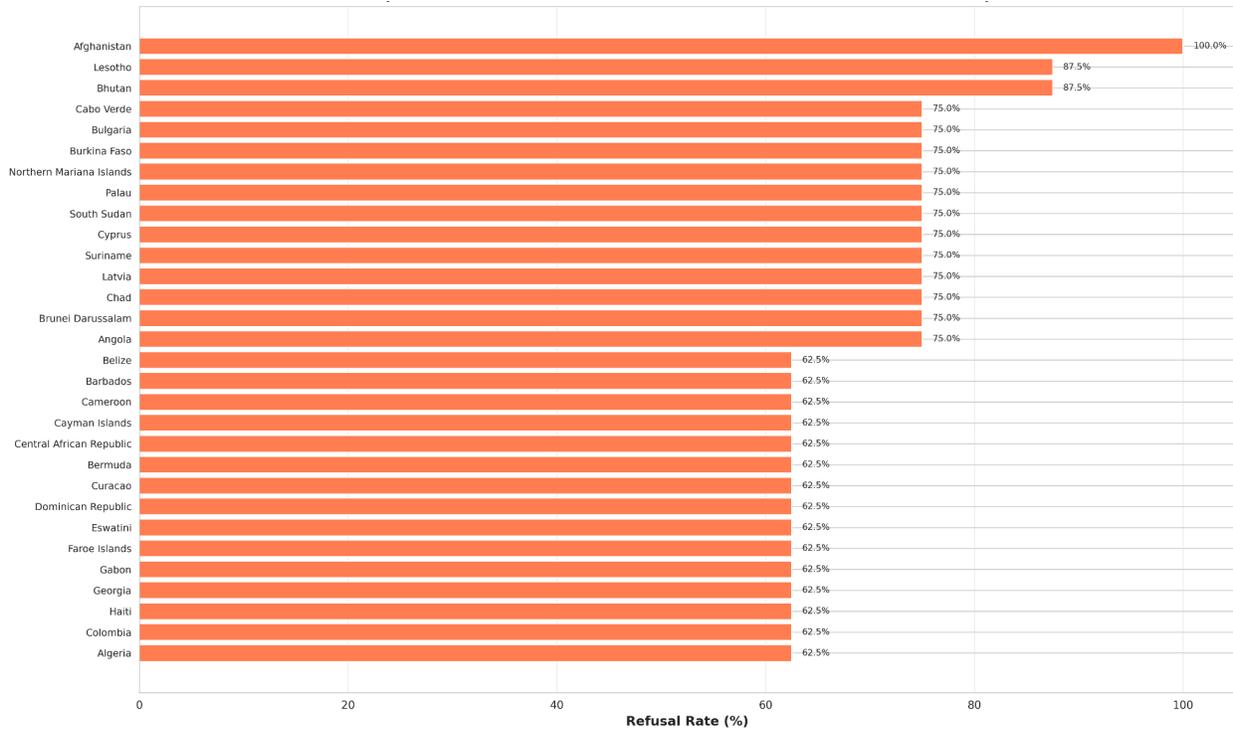

**Table 2: Top 30 Countries by AI Refusal Rate by Region**

| **Southeast Asia and Oceania** | Brunei Darussalam, Northern Mariana Islands, Palau |
|---|---|
| **Europe & Central Asia** | Bulgaria, Cyprus, Faroe Islands, Georgia, Latvia |
| **Latin America & Caribbean** | Barbados, Belize, Cayman Islands, Colombia, Curacao, Dominican Republic, Haiti, Suriname |
| **Middle East & North Africa** | Algeria |
| **North America** | Bermuda |
| **South Asia** | Afghanistan, Bhutan |
| **Sub-Saharan Africa** | Angola, Burkina Faso, Cabo Verde, Cameroon, Chad, Eswatini, Gabon, Lesotho, South Sudan, Central African |

| | Republic |
|---|---|

*Finding 2: Distribution of Countries with the Highest Knowledge Rates*

Graph 3 shows the knowledge rate heatmap of the top 20 countries across all eight metrics. As we discussed earlier, 11.4% of all responses are the provision of number/fact, which are considered factual because the type of answer is quantitative-based rather than empirically verified. Graph 3 demonstrates a list of 20 countries, where the Llama-3 8B model is most likely to provide factual responses rather than refusals, misunderstanding/correction or other responses. For each query corresponding to each metric, a factual response is considered 100% knowledge rate and all refusals, misunderstandings/correction or other responses are seen 0% knowledge rate. Graph 4 presents these top 20 countries by rates of factual answers. Graph 4 shares the same findings as Graph 3; yet, the former is more reader-friendly for readers to easily see the percentage of factual answers for these countries.

Table 3 shows the top 20 countries by knowledge rate by region. We can see there are some Western powers, such as Australia, Spain and Austria. There are additional high-income countries such as Saudi Arabia and Antigua and Barbuda. Surprisingly, there are also lower-middle-income countries, such as Laos from Southeast Asia and Angola from Sub-Saharan Africa, alongside Madagascar as a low-income country.

**Graph 3: Knowledge Rate Heatmap: Top 20 Countries x All Metrics**

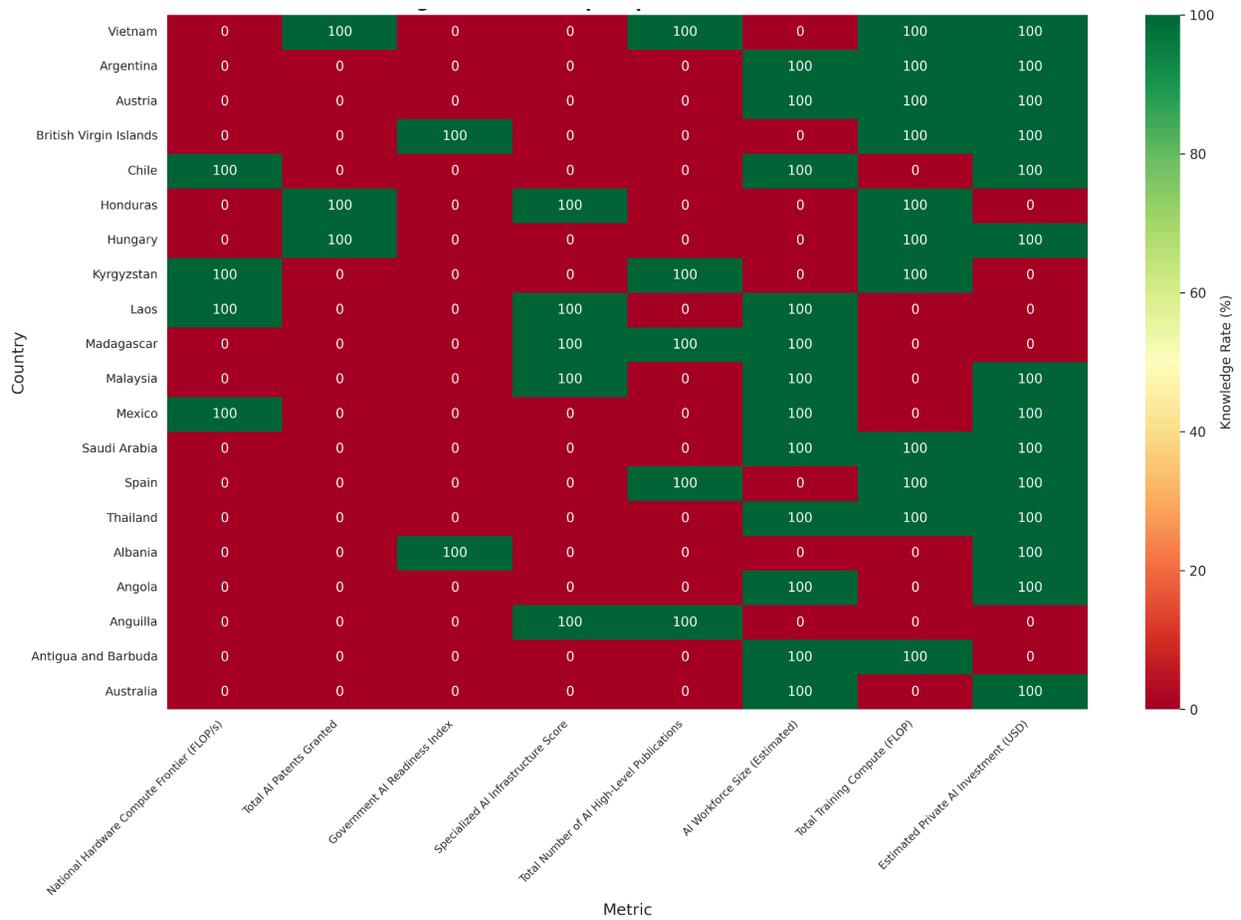

Graph 4: Top 20 Countries: Where AI Provides Most Factual Answers

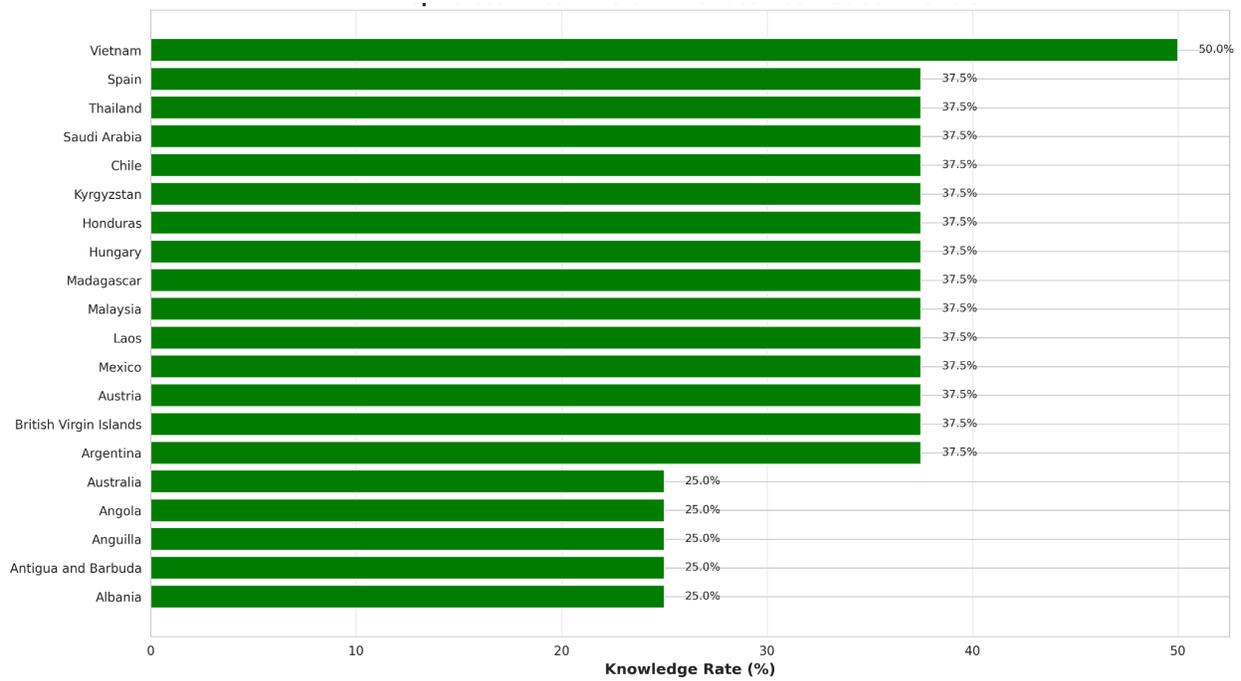

**Table 3: Top 20 Countries by Knowledge Rate by Region**

| **Southeast Asia and Oceania** | Australia, Laos, Malaysia, Thailand, Vietnam |
|---|---|
| **Europe & Central Asia** | Albania, Austria, Hungary, Kyrgyzstan, Spain |
| **Latin America & Caribbean** | Anguilla, Antigua and Barbuda, Argentina, British Virgin Islands, Chile, Honduras, Mexico |
| **Middle East & North Africa** | Saudi Arabia |
| **Sub-Saharan Africa** | Angola, Madagascar |

*Finding 3: Metric Difficulty*

Graph 5 illustrates the knowledge rate and refusal rate by metric among all 1704 queries. We can see that for each metric we measure, the refusal rate is higher than the knowledge rate. The metric measuring private AI investment (USD, estimated) has over 25% of the knowledge rate and over 40% of the refusal rate. For all other metrics, the gaps between the corresponding knowledge rates and refusal rates are much larger. Especially, for the government AI readiness index, total AI patents granted and national hardware compute frontier (measured in FLOP/s), we see there are only about 5% of the knowledge rates, but between almost 50% and over 60% of refusal rates.

**Graph 5: Metric Difficulty (in Terms of Knowledge Rate and Refusal Rate) by Metric**

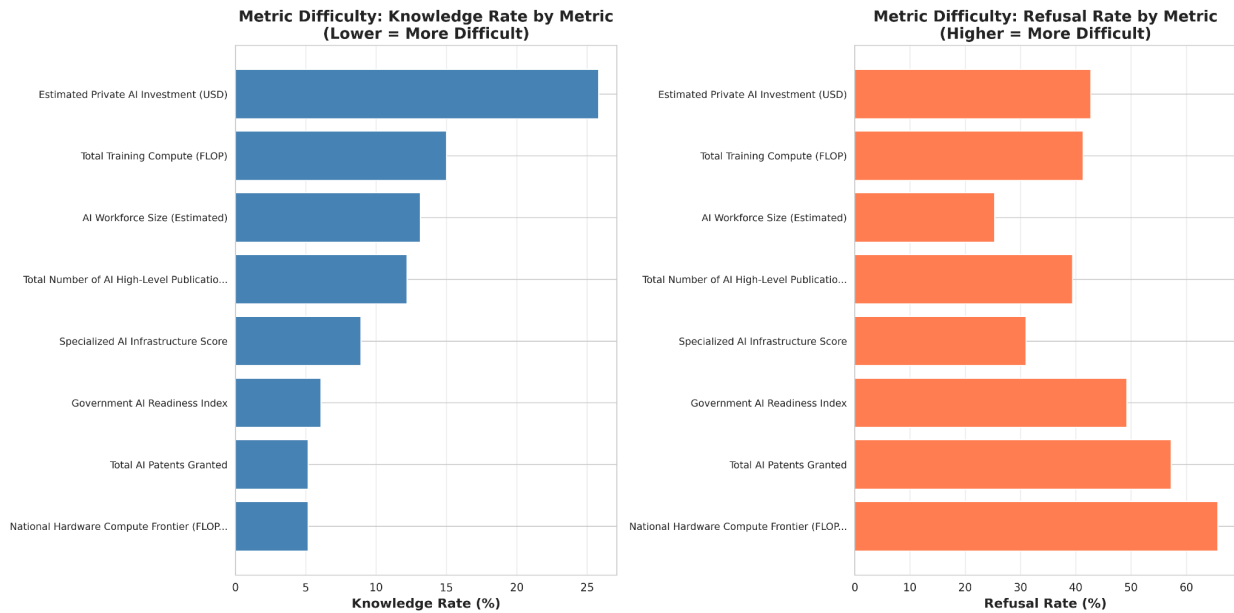

# 5. Discussion and Limitations

The results of this audit demonstrate the notable digital barriers in today's foundation models. For the Llama-3 8B model, only 11.4% of all query answers are considered factual responses. Also, these factual responses are unverified. In this exploratory project, unlike the actual Phases 3-4 of my GAID Project, I did not verify whether the numbers or facts provided by Llama in these 11.4% of the answers align with the metric values from the GAID dataset. Also, Llama was far more likely to provide refusal rather than factual responses in all metrics that I evaluated.

Moreover, the empirical findings of this exploratory project support existing arguments on the presence of geographical hallucinations. Countries and territories from the Global South are disproportionately more likely to be victimsed by knowledge gaps in these foundation models, where those from Sub-Saharan Africa, Latin America & the Caribbean, Oceanian island countries and South Asia are more vulnerable to knowledge refusals relative to knowledge delivery. Here, Latin America & the Caribbean and especially Sub-Saharan Africa and South Asia are widely recognised as the underdeveloped regions globally. In addition, Oceanian island countries are usually small states that are geoeconomically and geopolitically negligible.

## Limitations

- **Methodological Constraints**: The "Provides Number/Fact" category identifies quantitative responses but does not verify their empirical validity. Given that the queries requested 2025 data, the Llama-3 8B model might not have full access to such recent data. Therefore, this explains why the Llama-3 8B dissatisfactorily performed in this stress-testing process. For my Phases 3-4 GAID Project, I will develop a far more comprehensive and robust global AI bias audit across all foundation AI models to provide more scientifically rigorous and insightful findings.
- **Scope Limitations**: This study exclusively evaluated the Llama-3 8B model. I did not address how other open-weight or proprietary models handle the same data gaps. Again, this limitation will be addressed in Phases 3-4 of the GAID Project.
- **Timeframe Constraints**: Due to the 2-day hackathon timeframe, this audit was restricted to eight specific technical governance metrics extracted from the GAID dataset. I did not address failure modes related to the aforementioned linguistic and cultural bias highlighted in existing research.

## Future Work

I already clearly highlighted how this exploratory project is only a pilot study of my Phases 3-4 GAID Project. The detailed action plan of Phases 3-4 that I am going to deliver in the next 12 months is presented in my public post on EA Form, LessWrong and AI in Society. For those who are interested in learning more about the action plan for Phases 3-4, please feel free to read it at

https://forum.effectivealtruism.org/posts/DvzguceNKaQcYPgwb/the-global-ai-dataset-gaid-project-from-closing-research.

# 6. Conclusion

The global audit of AI technical governance awareness illustrates disparities in how LLMs provide data across regions and income levels. While LLMs are more likely to have data available for contexts related to high-income countries, especially Western democratic countries, they exhibit higher ignorance rates when answering prompts in the Global South. Therefore, international governance, policymakers and policy-focused scholars increasingly relying on AI agents for professional work may compound AI-driven systemic bias and, therefore, digital colonisation in AI-facilitated decision-making.

The research suggests that current AI models (at least the Llama-3 8B model) have yet to be maturely developed enough to serve as reliable tools for global technical governance. The digital barriers and gaps identified in this paper must be addressed through more inclusive data representation in model training and more transparent alignment processes to ensure that AI benefits—including safety, fairness and readiness—are accessible to all countries, regardless of their geographical location or income classification.

## Code and Data

*Include links if applicable. If your project doesn't involve code (e.g., policy analysis) or if there are info-hazard considerations, note that here.*

- *Code repository*:
  *[https://github.com/newlivehung123123/technical_governance_challenge_2026/tree/main/code_folder]*
- *Data/Datasets*:
  *[https://github.com/newlivehung123123/technical_governance_challenge_2026/blob/main/census_audit_results.csv]*
- *README*:
  *[https://github.com/newlivehung123123/technical_governance_challenge_2026/blob/main/README.md]*

## Author Contributions (optional)

*J.H. designed this exploratory project. He was solely responsible for data collection, analysis and the development of this manuscript.*

## LLM Usage Statement

*I asked Gemini Pro 3 for advice when I encountered technical difficulties in coding. Also, I used Gemini Pro 3 to help draft sections, followed by manual revisions.*